# Companding algorithm for the detection of malignant lesions in HDR CT mandibular images

Yuval Tamir, Hedva Spitzer, Noam Yarom, Yuval Barkan, Silvina Friedlander Barenboim and Alex Dobriyan

*Abstract*— Compressing and expanding (companding) HDR computerized tomography (CT) images to a single LDR image is still a relevant challenge. Besides the general need for simplification of the window setting method for the purpose of diagnosis, there are specific clinical needs for the resection of malignant lesions in the mandible, for example, where the lesions may exist in both the soft tissue and the bone. A successful medical algorithm has to take into account the requirement to variously expose the same range of specific gray levels, when they are presented by different body tissues. As a solution, we propose an adaptive multi-scale contrast companding (AMCC) algorithm that is implemented by using soft threshold separation to two channels (bone and soft tissue). This separation is determined by the HU intensity values (that are already computed in the algorithm) in the specific resolution that is best fitted to the teeth. Each channel contains different set of parameters. The AMCC algorithm successfully and adaptively compands a large variety of mandibular CT HDR images as well as natural images. Two collaborating physicians who evaluated tumor boundaries, by the "single presentation" method, reported that 92% of the algorithm output images were at least as useful for diagnosis as those in the window method, while 50% of the algorithm output images were better. When each slice was evaluated simultaneously, by the window setting method (bone and soft tissue) and algorithm output images, 93% of the evaluations declared a preference for the output images of the algorithm. We describe here a low-cost method for companding the HDR images with the ability to facilitate resections of mandibular lesions by providing optimal boundary definition.

*Index Terms*— Companding, HDR images, CT, Adaptive contrast enhancement, Mandible

## I. INTRODUCTION

Many algorithms for compressing and expanding (companding) LDR and HDR images have been developed during the last decade [1]–[3]. However, the need for companding medical images, such as CT scans, is still an open challenge.

Computerized Tomography (CT) images present the X-ray radiation attenuation coefficients of the body's tissues. Each pixel can range from -1000HU (air medium) to +1000HU (bones) and even higher [4], making them high dynamic range images. In contrast, the image screens have a low dynamic range (256 gray levels or ~900 gray levels for a radiological screen). Consequently, the HDR images are displayed using the ''window setting'' technique [4], [5]. In this common diagnostic procedure, a single CT slice is viewed several times (e.g. soft tissues window, lung window, liver window, and bone window) [4]. The window technique is also required and used for medical care and diagnosis of lesions in the mandible.

While the window setting technique can provide an appropriate diagnosis, this method has many drawbacks including the inability to simultaneously review the whole picture and the consequent time needed to complete the diagnosis[6], [7]. These drawbacks are particularly relevant beyond the need in hospitals in situations such as trauma [8]–[10] or combat [11], where the radiologists' diagnosis commonly creates a bottleneck. An additional crucial area is represented by the needs of a surgeon faced with pathology related to more than one type of body tissue, e.g. injuries in the chest area (bone, soft, and lung tissues) and the mandible (bone and soft tissues).

## Previous studies

A number of algorithms have been developed to compress the high dynamic images and some have been tested for the ability to compress HDR CT images, in order to enable the full dynamic range to be visualized in a single window. Most of the earlier algorithms for HDR CT images were global algorithms (belonging to the Tone reproduction curve (TRC) method [3]), which were based on the Histogram Equalization (HE) technique and its variations [12]–[15]. Later studies presented local compressing algorithms (belonging to the tone reproduction operator(TRO) method), such as the Adaptive Histogram Equalization (AHE) [12] and Contrast Limited Adaptive Histogram Equalization (CLAHE) [13]. These

Yuval Tamir is a member of the Electrical Engineering School, Tel-Aviv University, Tel-Aviv, Israel (e-mail: yuvaltamir87@gmail.com).
Hedva Spitzer is a member of the Electrical Engineering School, Tel-Aviv University, Tel-Aviv, Israel (e-mail: hedva@tauex.tau.ac.il).
Yuval Barkan is a member of the Biomedical engineering department and the Electrical Engineering School, Tel-Aviv University, Tel-Aviv, Israel
Noam Yarom is the head of the oral medicine unit, "Sheba" medical center, Tel-Hashomer, Israel. School of Dental Medicine, Tel Aviv University, Tel Aviv, Israel

Silvina Friedlander Barenboim is a member of the Oral Medicine Unit, "Sheba" Medical Center, Tel-Hashomer, Israel. School of Dental Medicine, Tel Aviv University, Tel Aviv, Israel.
Alex Dobriyan is the deputy chairman and residency director in the department of oral maxillofacial surgery, "Sheba" medical center, Tel Hashomer, affiliated with Sackler. Faculty of medicine, Tel Aviv university, Tel Aviv Israel.

algorithms were also tested on HDR CT images. The local algorithms successfully achieved some enhancement, mainly of lung tissue. A more recent variation of the AHE method is represented by Multi-scale Adaptive Histogram Equalization (MAHE) [16]. This method, which applies the AHE technique to the wavelet decomposition of the image [17] improved the success with lung tissue, but as reported: "Diagnostic accuracy, however, was insufficient in this pilot study to allow recommendation of MAHE as a replacement for conventional window display" [16].

Additional approaches with variations of Adaptive Contrast Enhancement (ACE) Algorithms [18] enhanced the contrast of the images through simple local statistical properties of each pixel's surroundings. Low frequency and high frequency components were assigned different values by the user in order to determine the level of enhancement.

A more recent study succeeded for the first time to compress and expand (compand) the full range of HDR CT images, such that the bone, soft, and lung tissues could be presented in a in a single window (LDR presentation). This BACCT (Biologically-based Algorithm for Companding CT images) algorithm employs a contrast adaptive method, which performs a curve-shifting mechanism that was inspired by the visual system [19].

Despite these advances, there remains an urgent clinical need to obtain a single image including both bone and soft tissues, for example when planning a mandibular resection. Over the last ten years or so, several studies have suggested the fusion of MRI soft tissue data together with the bone CT window (or CBCT image) [20]–[25], as a solution. Franz and his colleagues proposed the use of such a fusion [24] in order to resect tumor lesions in temporal bone. They used an existing navigation unit Stealth Station TM, which is based on improved patient registration software (CT and MRI). A later study performed this fusion with other software (Mimics 10.01 software Materialise, Leuven, Belgium) [23]. The fusion of CT and MRI was applied for the first time to mandibular resections, using Photoshop [20]. Different software has also be employed to provide an improved [25] but still not fully automatic fused image that can be used for the presentation of 3D virtual surgical planning [26].

A recent review [22] concluded that there is still insufficient knowledge regarding the degree of accuracy or clinical usage of CT-MRI fusion in the temporomandibular joints. The only algorithm that has attempted to tackle the fusion problem of the MRI and CT/CBCT images [21] suggested a non-subsampled shearlet transform (NSST) method. This is an advanced wavelet transform method that addresses the edges problem. The results were clinically evaluated using engineering methods.

Yet another approach suggested merging the different common HDR CT windows into a single image [27], [28]. The results of the RADIO (Relative Attenuation-Dependent Image Overlay) algorithm [27], [28] yielded blended images, which actually contain the information from several windows, but appear to be less detailed than each window presented separately. In addition, a possible flaw of such an approach could be that not all the diagnostic information necessarily exists in the different fixed windows (e.g. bone marrow). This problem is also relevant to our mandibular CT images, where both the teeth and bone tissue of the mandible often need to be exposed differently (not necessarily in the same fixed window).

A recent algorithm attempted to compress HDR images including CT images. This was done by a spatially weighted histogram equalization [29] method designed to develop a mixture of global and local tone mapping operators. Histogram equalization was used for the global part with a weighted neighborhood contribution for the local part. The method succeeded in compressing the CT HDR image to a single LDR image, but the challenge to compress CT HDR to achieve and better the resolution of a designated window, is still open.

We believe that a good algorithm that is able to compand the HDR CT windows to a single window could be efficient, for compression and enhancement of all body tissues. An additional issue in medical images that has yet to be addressed, is the requirement for different exposures of the same range of specific gray levels when they are presented in different body organs. For example, the details in teeth and bone should be very sharp and clear, whereas the same gray levels can appear in the soft tissue or in malignant tissue, but have to be exposed to reveal the coarse details.

These capabilities are necessary in order to visualize and detect abnormalities in different types of tissues (e.g. bone and soft tissue). For example, although the two tissue types may share the same range of gray levels, a metastasis can be visualized as a lump (coarse resolution), whereas clinical findings in teeth (e.g. vertical root fractures) require the exposure of small details (high resolution). To our knowledge, none of the existing algorithms can provide a solution for this issue and no algorithm has yet succeeded in efficiently companding HDR medical CT images. Here we describe an algorithm that can contend with these real-life challenges.

## II. PROPOSED ALGORITHM

### A. General overview

Our algorithm is inspired by the visual adaptation mechanism that enables us to distinguish fine details in different scales. The visual system can distinguish between an object and its surroundings as a result of the contrast strength of the object and its contrast context [30], [31]. Contrast and luminance, are modulated in such a way to increase the contrasts between an area and its surroundings. For example, in the case of a central area surrounded by an area with a weaker contrast, the central contrast will be strengthened [30], [32].

### B. The rationale behind the algorithm

Since the algorithm is basically designed to compand (compress and expand) the high dynamic range, we expect it to both reduce the dynamic range, on the one hand, and enhance the small details on the other. In addition, since the main goal of our algorithm is to compand HDR CT images to a single window, we need an additional process, which enables us to broaden the soft tissue narrow HU range (-200 to +300 HU). This is achieved through a pre-processing stage as described in the model section.

The inspiration for the main idea of the model came from consideration of the adaptation mechanisms in the visual system that perform "curve shifting" [33], [34], [35] in order to reduce the dynamic range or chromatic illumination. Here we refer to the adaptation of the contrast domain, rather than the intensity domain. Consequently, we chose to perform the adaptation through determining the "saturation" factor (Naka Ruston), and use a function of the contrast instead of a constant value, in the function of Naka Rushton response. The adaptation in this model is performed by changing the curviness of the Naka Rushton [36] according to the contrast differences between the current area and the neighboring region. The purpose of the curviness modulation in this model is to achieve an exposure of small details in areas of low contrasts by applying higher gains and to compress the high dynamic range by reaching saturation addictively. Because of the challenges posed by different spatial resolutions, the model is run on multi-resolution spatial scales. The adaptation is obtained through a multi-resolution texture modulation, which was suggested previously to be responsible for contrast adaptation mechanisms of the visual system [31].

In order to compand the HDR CT image into a single window that contains all the information of each separate window, we need to add additional components which are not required for companding natural HDR images. The first one relates to the requirement to expand the dynamic range of the soft tissue [19] & [37]. An even more difficult challenge is the need to expand various organs differently, even though they share the same gray level intensities. For example, the intensity level of the bone marrow may be similar to that of soft tissue, but if fine details are required (as in the teeth), the exposure needed is outside the usual range used for soft tissue. We suggest here a simple solution that does not require any segmentation operation. The idea is to use different algorithm parameters to separate the proposed algorithm to different channels, according to the level of intensity at a coarser spatial resolution, which for example is appropriate for the size of the teeth.

Our multi-scale adaptive contrast companding (MACC) algorithm is composed of 3 main stages where the first procedure is a pre-processing stage called soft tissue enhancement. This stage provides an enhancement in soft tissue window contrast. This is followed by a companding stage, using a compression and expanding (companding) algorithm, which includes our adaptation mechanism at different contrast resolutions, but with two separate pathways for bone and soft tissue. The third and final stage, is pyramid collapse, which enables the transition from multi-resolution contrast images with new adapted contrasts back to the two-dimensions intensity image with the finest resolution. The algorithm flow chart with all the stages is presented in Fig 1. The titles and sub-titles of the algorithm stages correspond to the titles in the flow chart blocks.

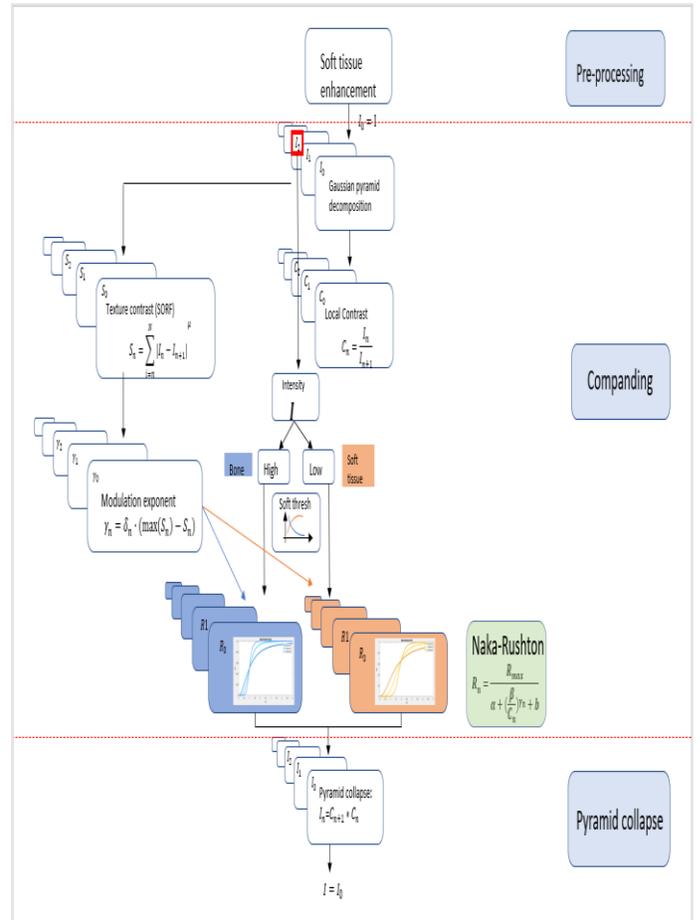

Fig. 1. Flow chart of our MACC algorithm presenting the building blocks and algorithm components for companding HDR images.

*C. Algorithm stages*

   *1. Pre-processing: Soft tissue enhancement*

The soft tissues window (ranging from about -200 to +300 HU) is the narrowest window where the range of CT slice in other tissues extends from -1000 to +3000 HU and higher. As a consequence of this narrow range, there are low levels of contrast in soft tissue. We applied a pre-processing enhancement component in order to expose the problematic low contrasts in the soft tissues in the original HDR image [19] & [37]. This module was developed previously and was inspired by the "window setting" technique. Here we apply it in our proposed algorithm as a pre-processing stage (Fig 1). This stage is first calculated by stretching the soft tissue range of the HDR CT image (Fig 3(b) in: [19]), Eq.(1) in [19].

$$W(x,y) = \begin{cases} \frac{4C_1(I(x,y)-V)(I_{max}-I(x,y))}{(I_{max}-V)^2}, & if \ I(x,y) \geq V \\ \frac{-4C_2(V-I(x,y))I(x,y)}{3V^2}, & if \ I(x,y) < V \end{cases} \quad (1.1.1)$$

Where $V$ determines the weight function W(x,y), $I(x,y)$ is the intensity value, $I_{max}$ is the maximal intensity value, and $C_1$ and $C_2$ are constants.

In addition, the pre-processing stage also includes a local contrast enhancement, but in this case this is achieved through testing whether the intensity of a specific pixel in the HU range differs significantly from those in the surrounding area [19]. For this purpose, two intensity HU ranges separated by a constant threshold (Figs 3&4 in: [19]) were selected, and then the contrast was evaluated for each range and treated as described in Eq.(2) in [19]:

$$I_{new}(x,y) = I(x,y) + W(x,y)\frac{G_{srnd}}{max(G_{srnd})} \quad (1.1.2)$$

Where $I_{new}(x,y)$ is the enhanced image and $G_{srnd}$ describes the local neighborhood.

## 2. HDR CT companding

The MACC algorithm compands HDR images into LDR images and enhances the contrasts, especially for the bone domain that was not included in the pre-processing stage (Fig. 1).

### a. Gaussian pyramid decomposition

A multi-resolution representation of the original image is produced by applying a Gaussian pyramid decomposition. Each resolution is the result of a Gaussian low pass filter and decimation of the previous resolution [38]:

$$B_{n+1} = Reduce(B_n) \quad (2.1.1)$$

Reduce operation is defined by:

$$B_{n+1}(i,j) = \sum_{m=-2}^{2}\sum_{n=-2}^{2} w(m,n) B_n(2i+m, 2j+n) \quad (2.1.2)$$

Where $w$ is a Gaussian probability distribution as commonly used in pyramid decomposition.
The finest resolution is set to be the original image:

$$B_0 = I \quad (2.1.3)$$

### b. Local contrast pyramid

The contrast is defined by dividing an image (or any other specific resolution in the pyramid) by a low pass filtered version. Due to the fact that the levels of the Gaussian pyramid are not the same size, the Gaussian pyramid level, n, is divided by the expansion of the next coarser level, n+1:

$$C_n = \frac{B_n}{Expand(B_{n+1})} = \frac{B_n}{E_n} \quad (2.2.1)$$

The expansion process is the opposite of the reduction process. This process is defined as an expansion of the image (or any other specific resolution in the pyramid) and an interpolation of the new pixel values between the previous versions [38]:

$$E_n(i,j) = 4\sum_{m=-2}^{2}\sum_{n=-2}^{2} w(m.n)B_{n+1}\left(\frac{i-m}{2}, \frac{j-n}{2}\right) \quad (2.2.2)$$

### c. Texture contrast (SORF)

In this work we used a texture contrast component that includes several contrast resolutions. This component can take into account edges that are not sharp and have no homogenous appearance. This is a non-linear component, which is inspired by a proposed model of a second order receptive field (SORF) in the visual system [31]. The SORF was built from a weighted sum of multi-resolution differences of Gaussians (DoG):

$$SORF(x,y) = \sum_{k=0}^{N}\frac{\iint |L_{SORF}^k(x,y)| \cdot W^k(x,y) \cdot dx \cdot dy}{\iint W^k(x,y) \cdot dx \cdot dy} \quad (2.3.1)$$

Where $L_{SORF}^k(x,y)$ is the difference between "center" and "surround" signals, which is in fact the difference of Gaussians (DoG). $W^k$ is a non-linear spatial weight function which is also dependent on $L_{SORF}^k(x,y)$. In other words, this SORF equation leads to a power law operation of each resolution before the summation stage, of the different contrast resolutions, $L_{SORF}^k(x,y)$ where $k$ is the resolution number index. Such a calculation enables us to enhance the dominant resolution at a specific texture without any detection of the prominent resolution.

Similarly, we define the "texture-contrast" as a weighted sum of the contrasts at different scales, and therefore define the "texture-contrast" (SORF) pyramid level, S, to be:

$$S_n = W_n \cdot |B_n - Expand(B_{n+1})|^\mu + (1 - W_n) \cdot Expand(S_{n+1}) \quad (2.3.2)$$

Where the last level, N, of the SORF pyramid is set to be:

$$S_N = |B_N - Expand(B_{N+1})|^\mu \quad (2.3.3)$$

For each resolution, the texture-contrast component, $S_n$, will be a weighted sum of the DoGs of all resolutions that are equal or coarser than n. We demonstrate the substitution of $S_{n+1}$ in Eq. 2.3.2:

$$S_n = W_n \cdot |B_n - Expand(B_{n+1})|^\mu + (1 - W_n) \cdot Expand\{W_{n+1} \cdot |B_{n+1} - Expand(B_{n+2})|^\mu + (1 - W_n) \cdot Expand(S_{n+2})\} \quad (2.3.4)$$

Substitution of coarser resolutions will continue until we reach the coarsest level $S_N$.

$W_n$ is a weight function that can be determined according to image type?
$\mu$ is the power law operation which enables us to further strengthen or weaken the contrasts.

### d. Modulation exponent

Our adaptation mechanism for contrast companding can increase the contrast if it is low and decrease the contrast where it is high. Moreover, the level of increase/decrease is dependent on the contrast in the neighborhood, as well as on the local contrast. In order to achieve this adaptive enhancement of the contrast, a modulation exponent, $\gamma$, is generated by applying a linear decrease function to the normalized SORF pyramid (2.3.2); Fig. 1:

$$\gamma_n = \delta_n \cdot (max(S_n - S_n) \quad (2.4.1)$$

where $\delta_n$ is a parameter which controls enhancement strength, since it determines the range of the available slopes for the Naka-Rushton response as explained below.

*Soft and bone tissues channels:*
Revealing and exposing both soft tissue and bone (teeth) details in the mandible, requires different organs to be treated differently even though they may share the same range of HU intensities. For example, the bone marrow (located spatially in the teeth) can share the same HU intensity values as in the soft tissue (orange channel for soft tissues and blue channel for bone in: Fig. 1). However, it is only necessary to expose the fine details in the teeth (e.g. to reveal vertical root fractures), while this enhancement is not actually recommended for soft tissue organs. Although segmentation could solve this problem, it could be considered an overkill for this kind of problem. To resolve this issue of two channels (bone and soft tissue), we propose an alternative simple method that uses the observations that the intensity values of the teeth are overall higher than those of soft tissues and that the teeth (at least for a specific jaw) have a fairly uniform size. To avoid potential problems due to consideration of partial areas in the teeth, such as the bone marrow, we selected the spatial resolution appropriate to the size of the teeth. The spatial resolution and high intensity values of teeth, allow us to differentiate between teeth and soft tissues and to define two channels each with a distinct set of parameters for the teeth and for the soft tissues.

To implement the channel separation, we modified the control of the enhancement strength [Eq. 2.4.1], such that the parameter $\delta_n$ becomes a function of the normalized intensity values ($\frac{B_m}{\max(B_m)}$ in Eq. 2.4.3) at the "teeth resolution" instead of a constant parameter [Eq. 2.4.2]. ST was chosen to be a Soft Threshold, [Eq. 2.4.3]. We applied the soft threshold as in [Eq. 2.4.4] as a complementary weight function for the two terms: bone and soft tissue channels. The first term is dominant for large normalized intensity values, which is more suitable for the bone channel, $\lambda_{bone,n}$, and the second term is dominant for small normalized intensity values, which is more suitable for the soft tissue channel, $\lambda_{soft,n}$.

$$\gamma_n = \delta(B_n) \cdot (\max(S_n) - S_n) \qquad (2.4.2)$$

$$ST = e^{-\frac{B_m}{\max(B_m)}} \qquad (2.4.3)$$

where $B_m$ is the intensity value of the specific "teeth resolution" and $m$ is resolution index of the "teeth resolution". $\max(B_m)$ is the maximum intensity value in the resolution that is compatible for the teeth.

$$\delta(B_n) = A \cdot (1 - ST) \cdot \lambda_{bone,n} + B \cdot ST \cdot \lambda_{soft,n} \qquad (2.4.4)$$

where A and B are constants, $\lambda_{soft,n}$ & $\lambda_{bone,n}$ are the values of pixel intensity at each resolution, n, of the soft and bone tissue channels, respectively. The separation in this equation [Eq. 2.4.4] is therefore probabilistic and not strict, i.e. when one term is more dominant the effect of the other one decreases but is not zero.

*e. Naka Rushton response*
The adaptive enhancement is performed on each level (resolution) of the contrast pyramid (Eq. 2.2.1) according to the exponent (Eq.2.4.1) of the Naka Rushton equation (Eq. 2.5.1). Each level of the response pyramid is defined to be:

$$R_{n_{modulated}} = \frac{R_{max}}{\alpha + \left(\frac{\beta}{C_n}\right)^{\gamma^n}} + b \qquad (2.5.1)$$

where $R_{max}$, $\alpha$, $\beta$ and $b$ are constant parameters.

The slope of the response increases as a function of $\gamma^n$ (Eq.2.4.1). Therefore, small changes in contrast will cause larger changes in the response, as shown in Fig. 2.

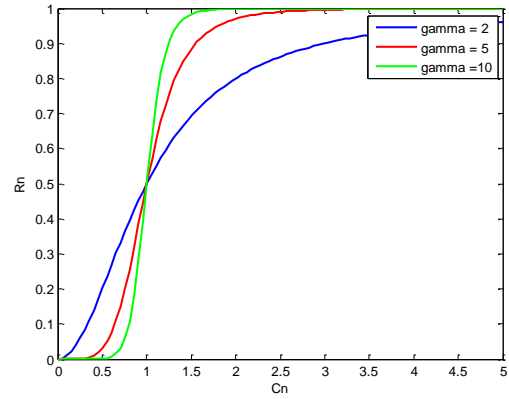

Fig. 2. Contrast adaptation mechanism for each contrast resolution. The figure presents the response curves as a function of the contrast level for the different gamma values, which reflect the level of adaptation. A higher gamma yields a steeper slope (green curve) for low values of contrast and reaches saturation at lower contrast values. All adaptation curves converge at $C_n = 1$, where there is no contrast between the two adjacent resolutions.

*3. Pyramid collapse*
The calculations described up to this point produce the modulated contrast pyramid levels, $\widehat{C_n}$. In order to apply the modulated image values to the resultant image, we perform a pyramid collapse of the modulated Gaussian pyramid levels, $\widehat{B_n}$. Each modulated Gaussian pyramid level is the result of the modulated contrast, $\widehat{C_n}$, by an expansion of the next coarser level. In fact, this is the inverse process of the contrast pyramid decomposition [38]:

$$\widehat{B_n} = \widehat{C_n} \cdot Expand(\widehat{B_{n+1}}) \qquad (3.1.1)$$

The coarsest (deepest) level of the modulated pyramid is set to be the deepest level of the original Gaussian pyramid (Eq. 2.1.1):

$$\widehat{B}_{n+1} = B_{n+1} \qquad (3.1.2)$$

The pyramid collapse progresses from the deepest level towards the finest level (resolution):

$$\hat{B}_N = \hat{C}_N \cdot Expand(\hat{B}_{N+1}) = \hat{C}_N \cdot Expand(B_{N+1})$$
$$\hat{B}_{N-1} = \hat{C}_N \cdot Expand(\hat{B}_N)$$
$$\ldots$$
$$\hat{B}_1 = \hat{C}_1 \cdot Expand(\hat{B}_2).$$
$$\hat{B}_0 = \hat{C}_0 \cdot Expand(\hat{B}_1) \quad (3.1.3)$$

The final modulated image is the finest resolution of the enhanced pyramid:

$$\hat{I} = \hat{B}_0 \quad (3.1.4)$$

Note that the original image will be restored if there is no modulation, i.e. $C_n = \hat{C}_n$,:

$$\hat{B}_N = \hat{C}_N \cdot Expand(\hat{B}_{N+1}) = C_N \cdot Expand(B_{N+1}) = B_N$$
$$\hat{B}_{N-1} = \hat{C}_{N-1} \cdot Expand(\hat{B}_N) = C_N \cdot Expand(B_N) = B_{N-1}$$
$$\ldots$$
$$\hat{B}_1 = \hat{C}_1 \cdot Expand(\hat{B}_2) = C_1 \cdot Expand(B_2) = B_1$$
$$\hat{B}_0 = \hat{C}_0 \cdot Expand(\hat{B}_1) = C_0 \cdot Expand(B_1) = B_0 \quad (3.1.5)$$

### III. METHODS

*Code and raw data:*

The algorithm in this study was implemented using the MATLAB software version R2016a. Nine full CT scans of patients diagnosed with malignancies involving the mandible were retrieved anonymously. All patients were diagnosed in the Department of Oral and Maxillofacial surgery at the Sheba Medical Center. The study was approved by the institutional review board. All scans (in DICOM format) were viewed by the Phillips DICOM viewer software.

*Pre-processing:*

All CT scans were first pre-processed in order to reduce the metal CT values, before applying the proposed algorithm. These clipped values are very high and have the tendency to dramatically extend the dynamic range, which then contains superfluous information. The results of the algorithm are stored in PNG format. This pre-processing phase is not required for natural images.

*Clinical image resources:*

The study was performed using a set of raw data images, which includes HDR CT images from patients. Four cuts for each patient; 3 axial images (at the level of the lower border of the mandible, mid-mandibular level, and at the level of the alveolar crest), and 1 mid-tumor coronal cut were selected by an experienced clinician (NY). Three images for each cut (a total of 108 images) were generated; the bone window image, soft tissue window image, and the new algorithm output image.

*Clinical test:*

In order to evaluate the algorithm results, the images were evaluated by two experienced physicians, one a maxillofacial radiologist (SFB), and the other an oral and maxillofacial surgeon (AD).

An online survey including all 108 images was prepared in order to compare the results of our algorithm to the conventional window setting. This survey was composed of two main parts: the first part was termed "Single presentation" and second was termed "simultaneous presentation". In the Single presentation, each of the 108 images, was randomly presented on a different page, together with two questions (per image). The first question was: "Please describe your ability to define the tumor's boundaries in the soft tissue". The second question was identical except that it referred to bone instead of to soft tissue. The physicians could choose to answer on a scale from 1 to 4, corresponding to well defined, somewhat defined, not so defined, and not at all defined. Each adjudicator evaluated all images using the same computer screen and under the same lighting conditions. The results with the algorithm outputs were compared to the physicians' evaluations from the standard windows method.

For the simultaneous presentation part, the physicians compared the 27 algorithm output images (taken from the patients' images at all axial cuts) with the corresponding two matching "window" images. Each web page in this clinical part of the questionnaire comprised 3 images simultaneously: algorithm output image, bone window image, and soft tissue image. In this part of the questionnaire, the physicians were required to choose the image which presented the best view of the tumor boundaries.

*Clinical results presentation:*

*Single presentation*: We divided the radiologist's rankings of each HDR CT image into 4 groups, according to the degree of success of the algorithm's performance:

The first group included the algorithm output LDR images that were ranked higher than both the matched "window" (bone or soft tissue) LDR images (green color/ Algorithm in Table 1). The second group included the algorithm output LDR images that were ranked higher than one of the matched "window" images, and equal to the other window image (orange color/ Algorithm ≥ soft/bone in Table 1).
The third group included the algorithm output LDR images that were ranked equally to both the corresponding "windows" images (yellow color/ Algorithm=soft & bone in Table 1).
The fourth group included the algorithm output LDR images that were ranked lower than one of the matching window images, and either higher or equal to the other window image (red color/ Algorithm ≤ soft/bone in Table 1). It should be noted that the final classification (Table 1) was calculated from the average of the ranking by both physicians.

*Simultaneous presentation*: For this part, we summed the physicians' answers across all 27 axial HDR images where they preferred the algorithm's output over the window setting LDR images.

### IV. ALGORITHM RESULTS

*Algorithm output results compared to the "windows settings" method:*

In order to evaluate the performance of our algorithm, we tested a series of CT images, of tumors that had spread to both bone and soft tissue in the mandible area. We chose CT slices where this type of tumors could be observed in both bone and soft tissue CT windows. The images presented below are examples that represent the algorithm's performance.

Figure 3 presents a CT slice of the axial mandible at a level where the teeth and bone are both exposed. This slice is presented in the bone and soft tissue windows, Fig.3a and Fig. 3b, respectively. Figure 3c presents the algorithm's results and demonstrates that both the soft and bone tissues (including the teeth and the jawbone) are exposed. Furthermore, it can be seen that the tumor area extends to both bone and soft tissues, blue arrow in Figs. 3a & 3b. There are two main problems with detecting the correct contours of the tumor by the windows method: (1) Neither window by itself provides the whole information about the spread of the tumor. For example, in the bone window the tumor is completely under-exposed in the soft tissue area, and vice versa in the soft tissue window. (2) The area of the tumor, in the bone tissue, for example, appears different in the two windows, even though it is actually the same tumor with the same spatial spread. More specifically, the appearance of the spatial spread does not reflect the actual spread in the window, which is not of the appropriate tissue type.

In contrast, Figure 3c successfully reveals the spread of the cancer in both the bone and soft tissue areas. In addition, the contours of the tumor are presented correctly in both tissue types.

In Figs. 4a-c, the lesion (blue arrow) cannot be seen in the bone window (Fig. 4a), while the tumor boundaries are partly obscured by the teeth in the soft tissue window (Fig. 4b), which has reached intensity saturation. In the algorithm output in Fig. 4c, the entire tumor area can be seen more clearly and can be better defined. In addition, all organs and body tissues can be seen well and identified in the single window of the algorithm output image (Fig. 4c). This is especially prominent in the palate exposure.

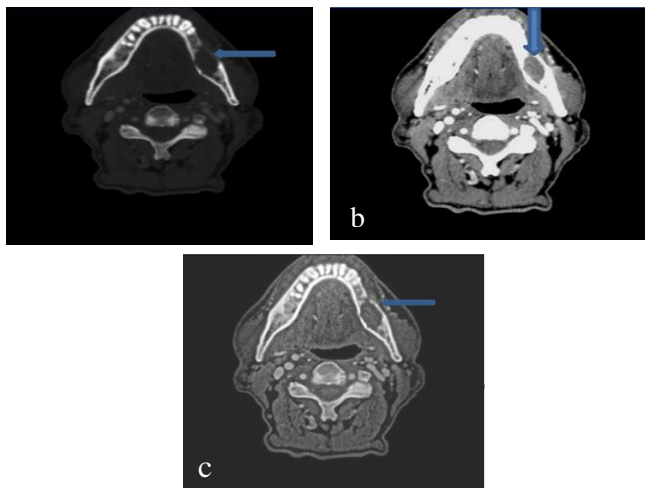

Fig. 3. Demonstration of our companding algorithm on mandibular CT image. Comparison of our MACC algorithm output image (c) to the conventional window method images ((a) Bone window (b) Soft tissue window). All three images contain the tumor lesion (blue arrow). It appears that the tumor boundaries are better defined in the algorithm output image than in the corresponding window images.

*Clinical results*

A summary of our clinical results is presented in Table 1. The results demonstrate that presentation of the whole dynamic range of the CT image in a single window does not impair the visibility of the required information. On the contrary, it actually causes the permits the tumor boundaries to be distinguished more clearly.

More specifically, in the "single presentation" method, 50% (a sum of the "green" and "orange" groups, Table 1) of the algorithm output images were considered to be improved in terms of visualizing the tumor boundaries in comparison to the window method. In addition, 92% of all the algorithm output images were ranked at least as high as the window method, while only 8% were ranked lower than one of the corresponding window images.

In the second part of the clinical test, where the three images (algorithm output, bone window, and soft tissue window) were presented simultaneously, in most cases, the two physicians considered that the best detection of the tumor contour was given by the algorithm output image (Table 1).

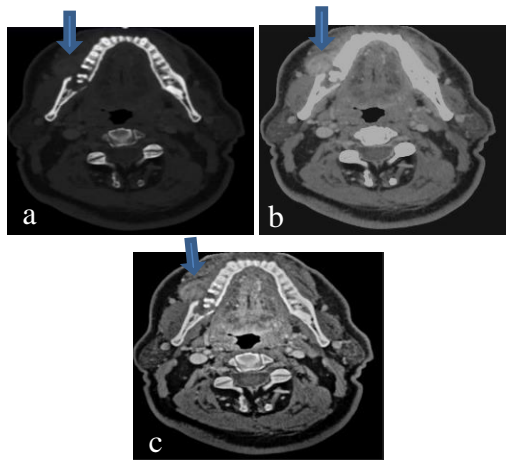

Fig. 4 - Demonstration of our companding algorithm on a mandibular CT image. Comparison of our MACC algorithm output image (c) to the conventional window method images ((a) Bone window (b) Soft tissue window). All three images contain the tumor lesion (blue arrow). It appears that the algorithm results contain more information than is presented in the soft tissue window.

The algorithm output images were considered superior to the windows method in visualizing the tumor boundaries in 100% (Physician 1) and 85% (Physician #2) of cases. This gave an average response of 93% of images judged in favor of the algorithm output.

TABLE I
SUMMARY OF PHYSICIAN EVALUATIONS

| Parameter | Physician1 | Physician2 | Average |
|---|---|---|---|
| Algorithm | 18.06% | 12.50% | 15% |
| Algorithm≥soft/bone | 29.17% | 41.67% | 35% |
| Algorithm=soft & bone | 43.06% | 38.89% | 41% |
| Algorithm≤soft/bone | 9.72% | 6.94% | 8% |
| Simultaneous presentation | 100% | 85% | 93% |

Summary of the responses of the two physicians to the clinical test questionnaire. The table presents the results of both the single and the simultaneous presentation evaluations. A set of 108 images were clinically evaluated in the single presentation section, while 27 triplets (algorithm output, bone window and soft tissue window images) images were assessed in the simultaneous presentation. The physicians' evaluations are presented here both separately and averaged across the two physicians. The group definitions are fully explained in the Methods

Although the two presentation methods were tested separately, they show the same trend where there is a clear preference for the images yielded by the algorithm.

*Natural Images:*

In order to assess the generic and robust nature of our algorithm we also evaluated the performance on natural HDR and LDR images This was done with a previous version of the algorithm that does not include the pre-processing stage [Eq. (1.1.1) & (1.1.2)] and the channel separation [Eq. (2.4.2) – (2.4.4)].

The performance of the algorithm was tested with the same set of parameters on dozens of diverse LDR and HDR images. We compared our MACC HDR results from several HDR images with those of several previously well-known algorithms [41]–[47] that also used multi-scale approaches. One of the most challenging situations in HDR imaging is compressing images that were taken under conditions of both indoor and outdoor illumination (e.g. bright sunshine entering a room).

## V. DISCUSSION

This study presents a generic and fully automatic algorithm that succeeds in companding the high dynamic range of CT images, as well as natural images, while presenting all the information in a single window. The algorithm can even slightly enhance the details in the images. Our algorithm successfully visualizes both soft tissue and bone tissue in a single window, to provide a clearer distinction of the correct borders of the malignant tumor.

This was achieved by adjusting the treatment of different organs even though they share the same range of HU intensities. Our method uses channel separation, without the need to perform high computational cost segmentations. This ability enables us to optimize resections by supplying more accurate information about tumor margins in critical organs such as the mandible.

The clinical preliminary results showed a significant statistical improvement (a-parametric test – Mann Whitney statistic) in defining tumor boundaries, through the algorithm output. The algorithm's performance has been assessed by observing images that contain lesions in both the soft and bone tissues in the mandible; Figs. 3-4. Clinical testing of the preliminary results showed that the diagnoses using the companded images were not inferior and were often better than the diagnoses using conventional window setting method, Table 1.

Our algorithm outputs show more accurate tumor boundaries in the mandible according to both our simultaneous and single presentation methods (Table 1).

A number of previous studies have also attempted to present all CT windows in a single window.
The first prominent algorithms to tackle this goal were the histogram equalizations and its variations [12]–[15] (including CLAHE). However these algorithms yielded insufficient contrast in the soft tissues as well as over-enhancement of image noises [7].

A later study of Fayad et. Al [7], which used a multi-scale adaptive histogram equalization method did achieve good results for lung tissue, but there was a lack of exposure in the soft and bone tissues.

While the algorithm of Cohen-Duwek and her colleagues [19] appeared to have succeeded in companding the whole CT image, the performance is improved by our current multi-scale approach and channel separation (with different level of details) of the soft and the bone tissues, such as required in mandibular CT.

The recent results of the blended CT windows studies showed success in presenting the 3 main windows of bone, soft tissue and lung, in a single window [27], [28]. In addition, the more recent study [28] showed impressive clinical results based on measuring the sensitivity and specificity, which yielded similar percentages of clinical performance to the conventional window method. However, when observing the studies' processed images [28], it appears that the different tissues (bone, soft and lung) are not as exposed as in the separate window images. This is more prominent for the bone tissue, where the bone marrow is under-exposed in all presented images, including window images.

Völgyes and his colleagues [29] tried recently to optimize local and global approaches in order to yield improved results in companding HDR CT images, without introducing artifacts such as the halo artifacts that most of the TRO tone mapping (local algorithms) suffer from. When comparing this article's output results to the different "windows" (original image), the results show insufficient contrast mainly in the bone marrow and in the soft tissue. In addition, according to the presented images it appears that the contrast in the lung tissue is inferior to that of the "lung window" (Dicom image file) and also to a previous study [19]. Volgyes et. al [29] compared their results to several previous algorithms by applying the successful tone mapping algorithms on CT HDR images. All the presented brain CT images appear to have insufficient exposure of the brain and bone tissues (Fig. 6 in: [29]). However, while the chest images looked better with the algorithms of [49], [50], [48], and [29], the separate windows were still superior to the compressed images. Our algorithm does not suffer from under-exposure in the bone and soft tissues, including exposure of bone marrows.

In addition, none of the previous algorithms presented results of companded images of the mandible, in which the bone marrow and the soft tissue need to be treated in a different manner. This need might be important for additional organs or areas in the human body. Therefore, a comparison with our data is difficult.

If until now we have compared our algorithm to previous versions with respect to image appearance after HDR compression it is also important to compare the performance of the different algorithms on clinical results.

Several previous studies have suggested algorithms and methods designed to present all CT windows of the HDR CT images of the mandible, in a single window. However only a few of them presented statistical clinical tests to evaluate these methods. One of these methods involves image fusion of the CT bone window with the MRI of the mandible. Qiu et al. [21] used several engineering objective evaluation metrics in order to evaluate their model and compare it to other algorithms. Among these metrics are: STD, SSIM, MI, SCD, $Q_0$, $Q_W$, $Q_E$, $Q^{AB/F}$ and VIFF. The authors showed that their shearlet transform algorithm yielded the best clinical results in 6 of the 9 suggested metrics tests when comparing to clinical results of older algorithms.

A more recent study suggested 3D fusion method of mandibular CT and MRI imaging and compared their clinical results with an historical cohort from 2009–2014 (CT based only) [25]. They showed better clinical results than the historical cohort. Their criterion for comparison was based on testing the patients after resection at different distances from the estimated cancerous contours. Their histopathology results revealed that all the resection planes in the bone were tumor free, whereas only 96.4% were tumor free in the historical cohort. Furthermore, the resection could be done with smaller margins as a result of their 3D fusion method.

The recent paper describing the "window blending" method suggested an improved decision threshold for the different body tissues and also supplied clinical results [28]. The authors reported improved sensitivity and specificity compared to clinical results obtained by the standard "window settings" (sensitivity of 82.7% in comparison to 81.6% and specificity of 93.1% in comparison to 90.5%).

While a number of methods for preliminary clinical evaluations have been described, there is as yet no consensus on how to best obtain preliminary results. This is also true for segmentation tasks. A further example of a clinical evaluation method in the mandible organ has been described by Rana and his colleagues [51].

It is difficult to compare different clinical tests across different methods, as to which performs best in detection of the malignant edges for resection, for example [21], [25], [28], [51]. Therefore, it is also difficult to compare these results to our clinical results (Table 1).

We did not expect that our proposed algorithm would achieve better tumor boundary ranking than the classical window setting. Moreover, our main goal was to obtain results that were not inferior to the classical window setting. However, according to the physicians' reports (Table 1) the algorithm actually performed better than the classical window setting. This trend was seen in the results of both clinical test methods: Single presentation and the simultaneous presentation methods (Methods section). Consequently, we can conclude that the algorithm has the potential to provide a more accurate presentation of the lesion boundaries for radiologists and for physicians who perform resections, mainly in sensitive areas such as the mandible.

In addition, due to the fact that in contrast to the several images required by the classical window settings, our output algorithm image presents all the information in one image, one can predict the saving of significant diagnosis time. This was not specifically tested in this study.

While our method requires further testing for clinical resection its preliminary success and low-cost relative to the very expensive fusion of MRI and CT [25] make it well worth further investigation.

VI. CONCLUSIONS

In summary, we describe a generic companding algorithm that we have developed for medical and natural images. Our results demonstrate that the algorithm enables a better diagnosis of the edges of a cancerous region in both soft tissues and bone in the same window. Consequently, it can facilitate an optimal resection that removes the malignant area while minimizing the volume of healthy vital area in the mandible that is removed. This approach has to be further tested and approved by more thorough clinical examinations, to support the preliminary results presented here. This algorithm has the potential to reveal additional clinical findings from a single image, even findings that have not been exposed before.